\documentclass[10pt]{amsart}

\usepackage{amsmath}
\usepackage{graphicx}
\usepackage{subfigure}
\usepackage{cite}

\title{Numerical study of a recent black hole lasing experiment}
\author{M. Tettamanti$^{1}$ \and S.L. Cacciatori$^{1,2}$ \and A. Parola$^{1}$ \and I. Carusotto$^{3}$}
\address{\noindent $^1$Dipartimento di Scienza e Alta Tecnologia, Universit\`a dell' Insubria, via Valleggio 11, 22100 Como, Italy \endgraf
$^2$INFN Sezione di Milano, via Celoria 16, 20133 Milano, Italy \endgraf
$^3$INO-CNR BEC Center and Dipartimento di Fisica, Universit\`a di Trento, I-38123 Povo, Italy}

\begin{document} 


\begin{abstract}
{We theoretically analyse a recent experiment reporting the observation of a self-amplifying Hawking radiation in a flowing atomic condensate~\cite{Stein2014}. We are able to accurately reproduce the experimental observations using a theoretical model based on the numerical solution of a mean-field Gross-Pitaevskii equation that does not include quantum fluctuations of the matter field.
In addition to confirming the black hole lasing mechanism, our results show that the underlying dynamical instability has a classical hydrodynamic origin and is triggered by a seed of deterministic nature, linked to the non-stationary of the process, rather than by thermal or zero-point fluctuations.}
\end{abstract}

\maketitle


\section{Introduction}

In 1974 S.Hawking demonstrated that black holes ``evaporate'' as a consequence of quantum effects~\cite{Hawking2}. That is, they emit particles in the form of thermal radiation at a temperature $T_H={\hbar \kappa}/({2 \pi k_B c})$ where $\kappa$ is the horizon's surface gravity, $\kappa={c^4}/({4GM})$ for a Schwarzschild black hole. Observing this stunning feature is not an easy task, due to the fact that the temperature depends on the inverse of the black hole mass; in fact, for the case of an astrophysical black hole of mass $M$, we have $T_{H} \simeq 10^{-7} \frac{M_{\odot}}{M}$ K which is well below the cosmic microwave background radiation temperature; therefore, a revelation of this phenomenon through an astrophysical observation is very unlikely. \\ On the other hand, the Hawking effect is a kinematical effect of quantum fields in curved spacetime and it only relies on the existence of a curved metric and a horizon. Thanks to these characteristics, it exists an alternative way to test Hawking's theories which was proposed by W. G. Unruh in 1981 \cite{Unruh} and is based on the mathematical analogy between the propagation of long wavelength sound waves in inhomogeneous and moving fluids and that of massless scalar fields in curved backgrounds. In particular, sound waves get trapped in regions of supersonic flow in the same way as light is trapped inside a gravitational black hole. This was the starting point to develop an analogy from gravitational black holes to acoustic ones. This way, in fact, one could seek to reproduce analogues of black holes in the laboratories and, pushing the analogy further to the quantum level, one could hope to observe effects of quantum field theory on curved backgrounds in acoustic analogue systems. This opened up new possibilities to experimentally test Hawking's predictions. In the past years many systems in different fields of physics have been explored towards the observation of this effect. Among the most promising ones, pulse propagation in non-linear optical media~\cite{Philbin,Belgiorno2010,Finazzi2}, quantum fluids of light~\cite{RMP,Gerace,Nguyen,BarAd,ICpropag}, atomic Bose-Einstein condensates of ultra cold atoms~\cite{BEC,Carusotto2008,Stein2010} and, restricting to a classical regime, surface waves on water~\cite{WeinfurtnerPRL2011,Parentani14,Rousseaux}. 

This Letter is devoted to a detailed study of a recent experiment performed by Jeff Steinhauer at Technion and claiming the observation of a self-amplifying Hawking radiation via the black-hole laser mechanism in a flowing atomic Bose-Einstein condensate \cite{Stein2014}. 
We make use of a numerical simulation to quantitatively reproduce the experimental data. Based on this theory, we discuss the implications of the experiments in the framework of the on-going quest for the experimental observation of a spontaneous Hawking emission from zero-point fluctuations.

\section{The 2014 Technion experiment}

The experiment described in \cite{Stein2014} creates a sonic black hole by means of a Bose-Einstein condensate accelerated to supersonic speed. The setup is realized with $^{87}$Rb atoms subject to a combination of magnetic and optical potentials: a magnetic field gradient and a laser beam constrict the condensate in the transverse directions to a tube-like volume, forcing the dynamics to be nearly one-dimensional, and provide a shallow confining trap potential along the longitudinal direction. Then a suitable, step-like longitudinal optical potential (called the ``waterfall'' potential) accelerates the atoms above the speed of sound within a finite region along the longitudinal direction, thus creating a pair of analogue black hole and white hole horizons: on one side of the region the phonons cannot exit while on the other side they cannot enter. 

\begin{figure}[h!]
\includegraphics[scale=0.59]{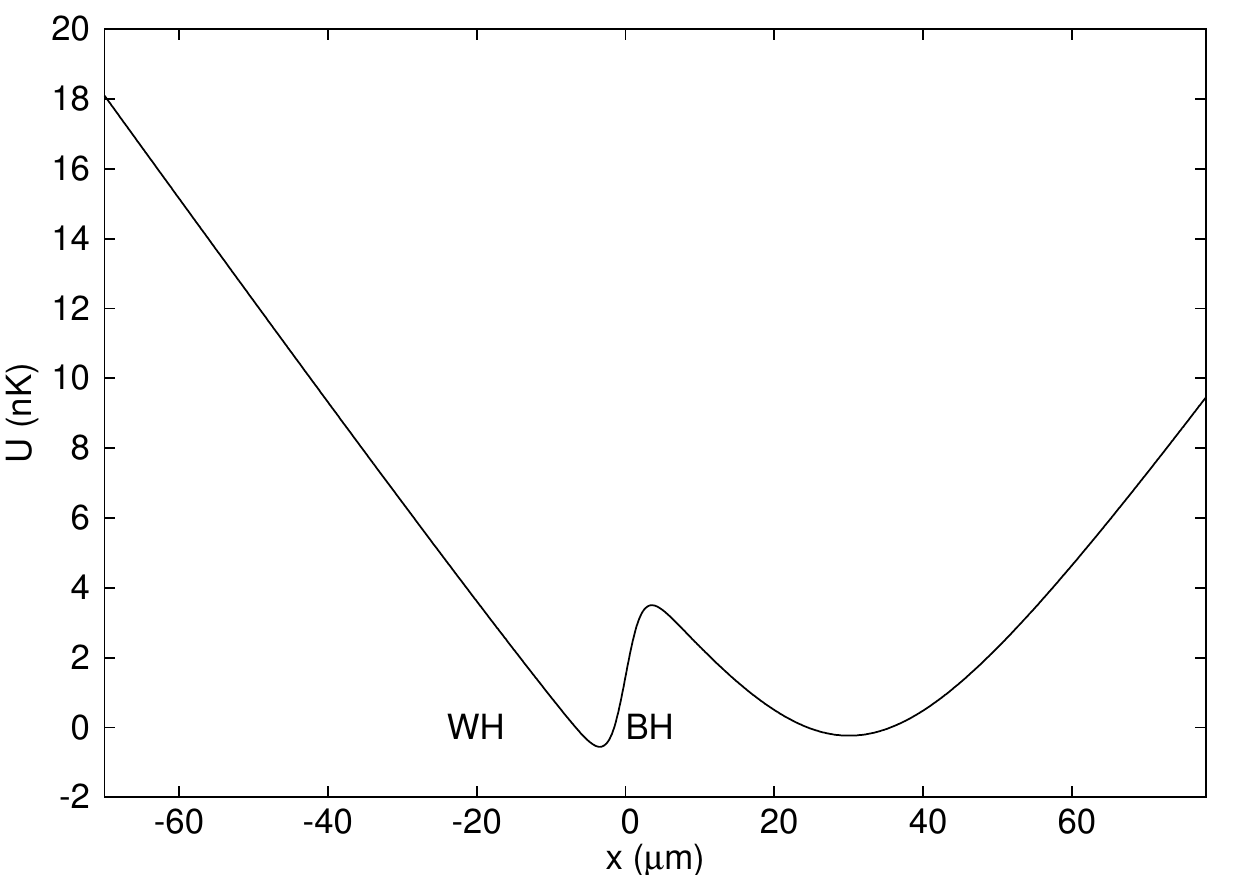}
\caption{Snapshot of the longitudinal potential used in our simulation (to be compared to Fig.~1c from\cite{Stein2014}) at time $t=20$ ms. WH and BH indicate the position of the inner white hole horizon and the outer black hole horizon respectively. In our numerical simulation, the condensate flows from the right to the left, which is Galilean-equivalent to the experimental configuration where the step-like waterfall potential is moved in the rightward direction.}
\label{potenziale}
\end{figure}

In Fig.~\ref{potenziale} we give a representation of our longitudinal potential, which was constructed in order to fit the experimental one. During the experiment, the confining part of the longitudinal potential remains at rest while the step-like waterfall potential is swept along the condensate at constant speed in the rightward direction; this way, on the right side of the step the condensate is slowly moving while on the left side it flows at a much faster supersonic speed with respect to the waterfall potential. The step is therefore responsible for the creation of the black hole horizon, which moves at a constant speed. Due to the shallow confining trap potential, the flow velocity far away on the left side of the step slowly decreases again until it gets again subsonic at the white hole horizon. 

In the experiment, the system is let evolve for 120 ms and the condensate density is imaged at seven instants of time with a phase contrast imaging technique. For each of these times, an ensemble of approximately 80 images is collected and then averaged to reduce noise. From the resulting density profile, the spatial profiles of the condensate velocity and of the speed of sound are finally extracted.


\section{The numerical model}

The starting point of our model is the time-dependent Gross-Pitaevskii equation (GPE) \cite{Stringari}, which represents a good approximation for the dynamics of a dilute Bose gas. In order to describe a nearly one-dimensional system, we follow Ref. \cite{Parola} 
and trace out the transverse degrees of freedom, so to obtain the so-called Non-Polynomial Schr\"{o}dinger Equation (NPSE) appropriate for a harmonic transverse confinement potential:
\begin{equation}
\label{gpe1d}
\begin{array}{l}
i \hbar \dfrac{\partial f}{\partial t} = \left( -\dfrac{\hbar^2}{2m}\dfrac{\partial^2 }{\partial z^2} + V + \dfrac{gN}{2\pi a_{\bot}^2} \dfrac{|f|^2}{\sqrt{1+2 a_s N |f|^2}} \right) f + \\ 
\quad \quad \quad + \dfrac{\hbar \omega_{\bot}}{2} \left( \dfrac{1}{\sqrt{1+2 a_s N |f|^2}}+ \sqrt{1+2 a_s N |f|^2} \right) f \, ,
\end{array}
\end{equation}
where $f(z,t)$ is the longitudinal part of the wavefunction and the transverse profile has been assumed to be Gaussian (for a system of weakly interacting bosons $g=\frac{4\pi\hbar^2a_s}{m}$, where $a_s$ is the bosons' s-wave scattering length). Here $a_{\bot}=\sqrt{{\hbar}/({m\omega_{\bot}})}$ is the transverse harmonic length, $\omega_{\bot}=965$ rad/s is the transverse trap's frequency, $N=4670$ the total number of bosons and $z$ the axis of motion. The longitudinal potential $V(z,t)$ is assumed to have the form:
\begin{equation}
\label{axialpotential}
V(z,t)=V_0\sqrt{(z-z_0+vt)^2+a^2}-\frac{V_s}{2}\left[1-\tanh\left(\dfrac{z}{\sigma}\right)\right]+c ,
\end{equation}
with parameters $V_0=0.3$ mK/m, $z_0=44.63\, \mu$m, $v=0.21$ mm/s, $a=19.44\, \mu$m, $V_s=6.4$ nK, $\sigma=2.33\,\mu$m and $c=-6.07$ nK chosen to fit the experimental ``waterfall'' potential described in \cite{Stein2014}. In particular, to better match the experiment, the step-like waterfall potential has been smoothed with a $\tanh(z)$ function. A snapshot of the resulting total potential at time $t=20$ ms is represented in Fig.~\ref{potenziale}.
Note that for numerical convenience reasons, we have preferred to move the trap in the leftward, negative-$z$ direction and to keep the waterfall potential at rest. This setting is anyway fully equivalent to the experimental one in~\cite{Stein2014} since the two configurations are related by a Galileian transformation. 

As an initial condition, we consider the ground state wavefunction that is obtained from an imaginary-time NPSE evolution in the presence of interactions and of the harmonic trapping but in the absence of the waterfall potential. More precisely, we start the evolution at $t=-50\,$ms when the step of the waterfall potential is still far to the left of the trap minimum and the condensate does not touch it. According to the above-mentioned Galilean transformation, the atoms are also given an initial velocity in the negative $z$ direction equal to the trap's velocity, so no spurious acceleration stage is introduced. 


\section{Results}

We proceed by numerically integrating eq.~(\ref{gpe1d}) and evaluating the axial density $n(z,t)=N|f(z,t)|^2$. As a first result, we verify that the configuration of a black hole-white hole pair is recovered by calculating the condensate and the sound velocity. Within the NPSE approach\cite{Parola}, the former is obtained by means of the one-dimensional continuity equation\footnote{Note that the result is qualitatively unchanged even if time steps as long as $20$\,ms are taken, as actually done in the experiment.} as:
\begin{equation}
 \label{v}
 v(z) = -\dfrac{1}{n(z)}\int^{z}_{-\infty}\dfrac{\partial n(z')}{\partial t}\,dz'.
 \end{equation}
while the latter, averaged over the transverse directions, is
\begin{equation}
c(z)=a_{\bot}\omega_{\bot}\sqrt{\dfrac{g'|f(z)|^2}{\sqrt{1+g'|f(z)|^2}}} \, ,
\end{equation}
where $g'=2a_sN$ is the effective one-dimensional interaction constant. 

\begin{figure}[h!]
\begin{center}
\subfigure[\label{velocitanew}]{\includegraphics[scale=0.55]{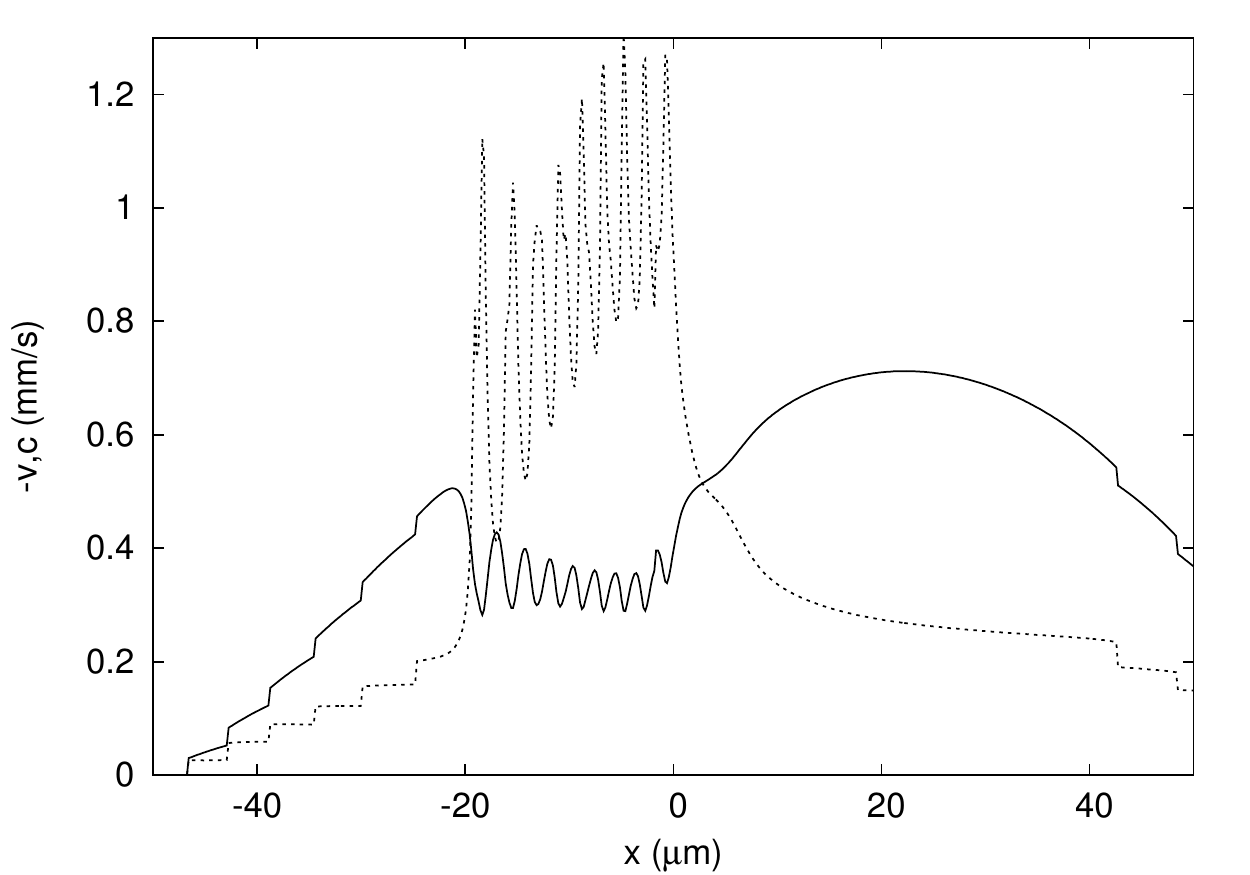}}
\subfigure[\label{velocitanewnoise}]{\includegraphics[scale=0.55]{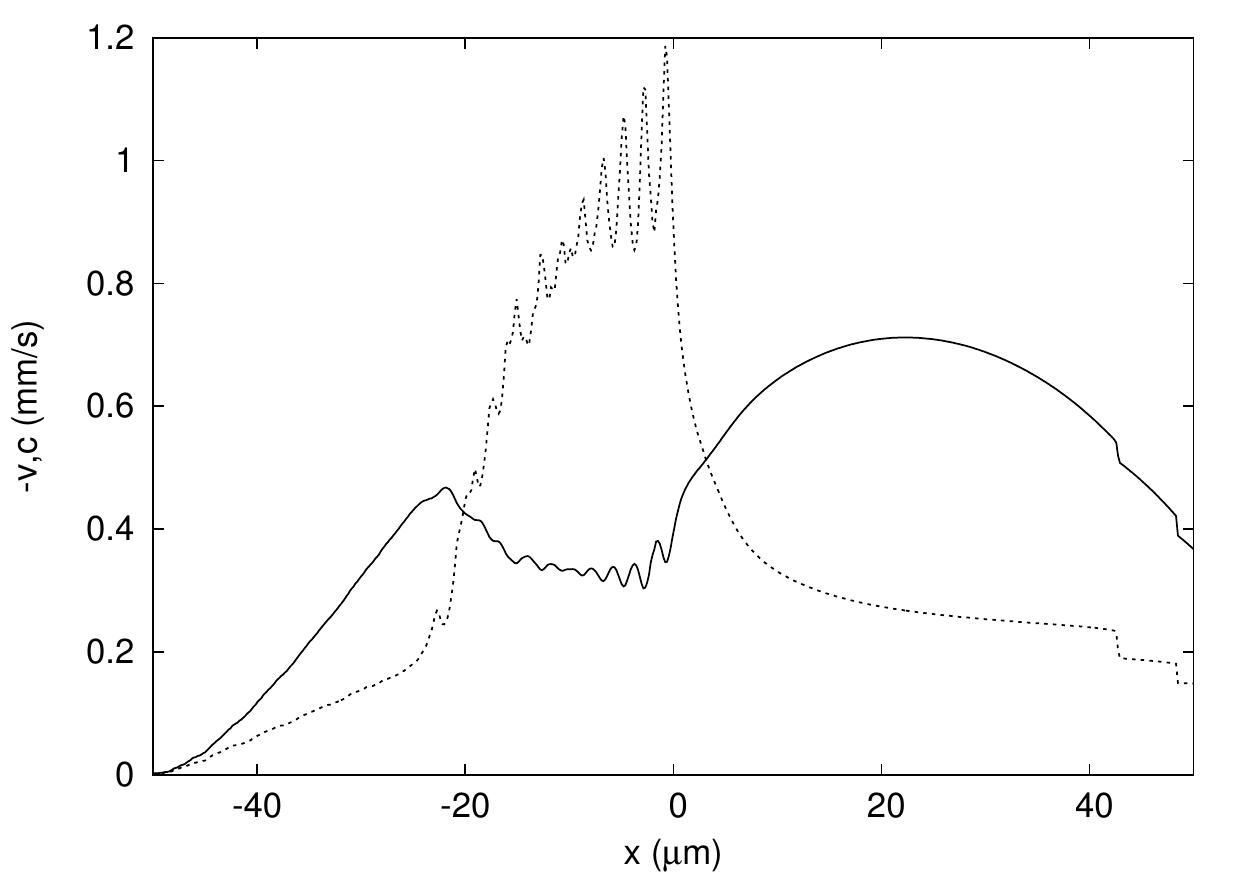}}
\end{center}
\caption{The dotted and the solid curves indicate $-v$ and $c$ respectively. The left panel Fig.~\ref{velocitanew} shows the prediction for the two velocities from eq.~(\ref{gpe1d}) with no additional noise. The right panel Fig.~\ref{velocitanewnoise} shows the same curves in the presence of fluctuations in the position of the white hole horizon. As it is discussed in the text, these are included by means of 10\% fluctuations on the intensity of the waterfall potential $V_s$. In both panels the curves are the temporal average over the 120~ms corresponding to Fig.~\ref{densityprofiles}. A well-defined supersonic region is clearly visible inside the two horizons in both cases.}
\label{velocita}
\end{figure}

The averages of these two quantities over the evolution interval are shown in Fig.~\ref{velocitanew}.
A pair of analogue black hole and white hole horizons is evident from the presence of a spatially limited supersonic region. It is important to notice that our results agree with the experimental observations shown in Fig.~3a of \cite{Stein2014} from a quantitative point of view also. Indeed, the sound velocity (solid curve) and the width of the lasing region (the region of supersonic flow) agree with the experimental results. The only minor discrepancy lies in the stronger contrast of the fringes shown by the speed of sound and, even more visibly, by the condensate velocity (dotted curve) in between the two horizons. As it is mentioned in the experimental article, there are some shot-to-shot fluctuations in the position of the white hole horizon. As the experimental results are obtained by averaging over many different realisations of the experiment, the decrease in the fringe contrast in the region between the two horizons is likely due to these fluctuations. A brief discussion of their origin will be given in the following of this work.

\begin{figure}[h!]
\begin{center}
\includegraphics[scale=1.1]{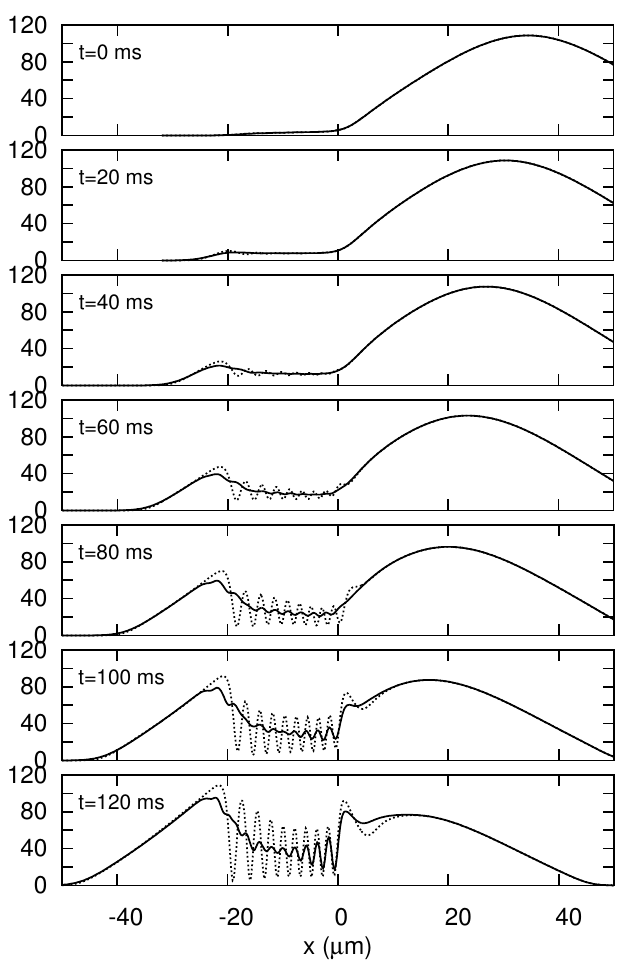}
\end{center}
\caption{The evolution of the condensate's density profile for 120 ms. The snapshot profiles shown are taken at 20 ms time intervals and the unit for the \textit{y}-axis is $\mu m ^{-1}$. The dotted curves show the density profile without any additional noise. The solid curves represent the average of the density profiles over 80 realizations when fluctuations in the position of the white hole horizon are included as done in Fig.~\ref{velocitanewnoise}.}
\label{densityprofiles}
\end{figure}

\section{The density profiles}

We turn our attention to the density profiles which are shown as dotted lines in Fig.~\ref{densityprofiles}: the agreement with the corresponding experimental profiles shown in Fig.~2j-p of \cite{Stein2014} is good both in the overall condensate density and in the width of the lasing region. Most importantly, the characteristic fringe pattern in the region between the two horizons is also quantitatively recovered and it is apparent how the amplitude of these fringes grows in time, as observed in the experimental data. In \cite{Stein2014} this was interpreted as a main signature of the so-called ``black hole laser'' effect \cite{blackholelasers,Parentani2}. 

When two horizons are present, sound waves which bounce back and forth in the region between the two horizons get amplified by the horizons at each reflection via a mode-conversion mechanism that often goes under the name of stimulated Hawking emission. The interference of these two counter propagating waves produces a growing fringe pattern in the region between the two horizons, which behaves as a laser cavity for sound waves. In the classification of instabilities proposed in~\cite{michel2013saturation,parentani4,Munoz,finazzi}, our configuration seem to correspond to a dynamical instability purely imaginary energy showing no oscillations.
From the presence of the self-amplifying negative energy waves between the horizons, it is then natural~\cite{Stein2014} to infer that a growing density modulation must be emitted from the region between the horizons into the outer condensate: even though this effect is partially obscured by the finite size of the actual condensate, a such density modulation can be spotted in the late time panels of Fig.~\ref{densityprofiles}, mostly in the $x>0$ region.

As compared to the experiment \cite{Stein2014}, it is important to note how the fringe pattern predicted by our simulations again shows a larger contrast. As already mentioned for the condensate velocity and the speed of sound, this is likely to be due to the experimental fluctuations in the position of the white hole horizon; a complete discussion of this effect, including some analysis of the fine structure of the fringe pattern, will be given in the following of this Letter.
Nevertheless, the key conclusion of our work so far is that the same interference pattern can be reproduced starting from a simulation based on the mean-field Gross-Pitaevskii description of the condensate. In our theory, the phonon field, whose zero-point fluctuations would be responsible for the spontaneous Hawking radiation, is fully classical. Yet the theory recovers the same interference pattern observed in the experiment, which indicates that the phenomenon observed is indeed due to a classical hydrodynamical instability effect.

\section{The seed}

It remains to be explained what is the initial seed for the dynamical instability leading to the eventual fast growth of the fringe pattern. The numerical simulations of Fig.~\ref{densityprofiles} again provide a neat answer: as it is particularly clear in the third panel from the top for $t=60$~ms, the wave pattern is seen to first appear at the white hole horizon and then to propagate towards the black hole horizon. This suggests that the most natural candidate for the seed is the Bogoliubov-Cerenkov effect \cite{bogoliubovcerenkov}, also called ``undulation'' in the analogue model literature~\cite{coutant2012hawking}: This is a linear process and is responsible for the appearance of a small static density modulation whenever a condensate flowing at supersonic speed encounters a weak obstacle, in our case the growing harmonic trap potential in the region around the white hole horizon. This initially small modulation gets then strongly amplified by the BH lasing instability. The deterministic nature of the Bogoliubov-Cerenkov emission process reflects in the fact that the fringes also have a deterministic character, with a well defined position and amplitude. In particular, this shows that there is no need of invoking non-linear effects to explain the non-vanishing average of the density modulation when starting from a stochastic, either quantum or thermal, seed~\cite{michel2013saturation}.

\begin{figure}[!ht]
\begin{center}
\includegraphics[scale=0.59]{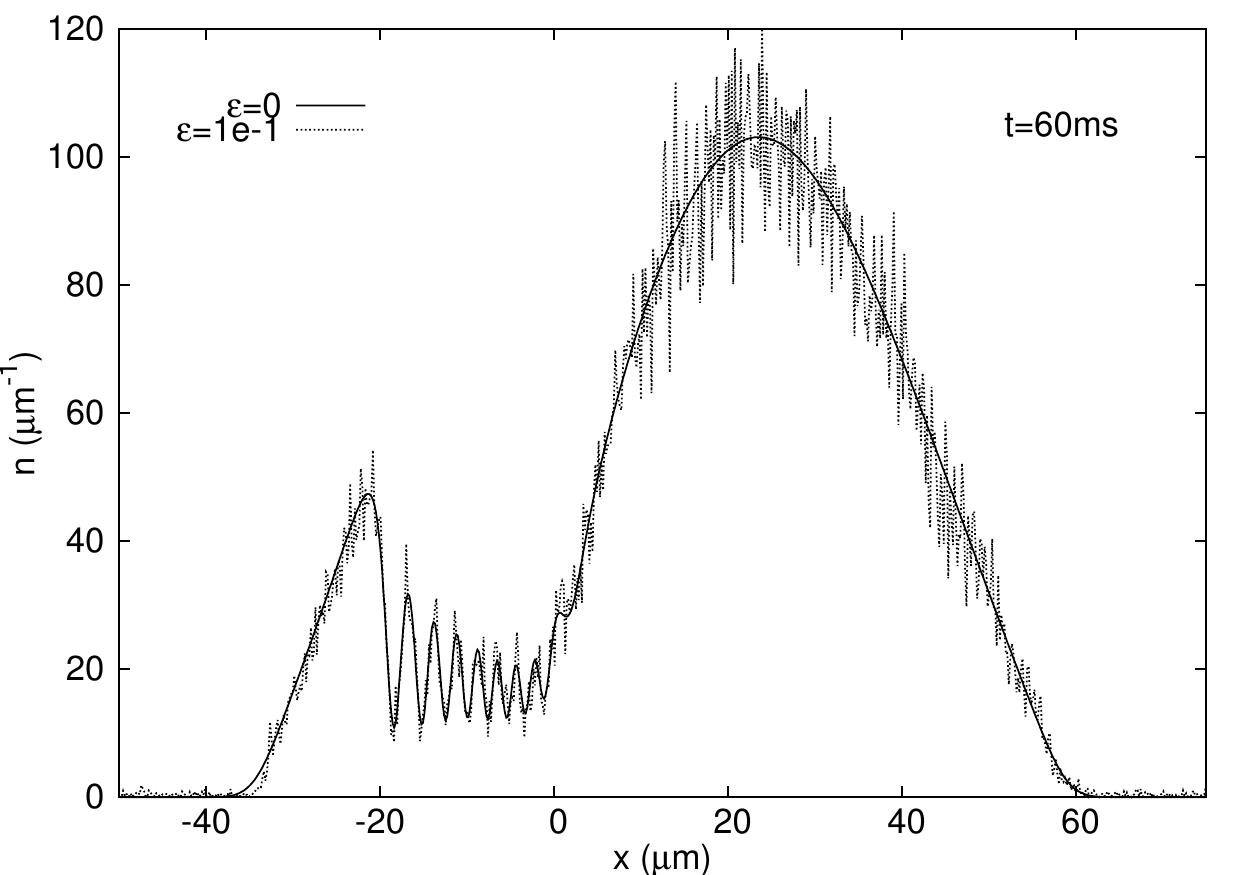}
\end{center}
\caption{The density profiles at $t=60$~ms. The solid curve shows the density profile when no noise is added while the dotted one is obtained by adding a random gaussian noise of intensity $\epsilon=10^{-1}$ to the initial condition.}
\label{noise80}
\end{figure}

To further illustrate the stability of the classical dynamics against additional fluctuation effects, we have repeated our simulations adding some noise to the interacting ground state used as an initial condition of the NPSE evolution,
\begin{equation}\label{noiseeq}
\bar f_0(z)=f_0(z)\,[1+\epsilon \eta(z)],
\end{equation}
where $\eta(z)$ is a random variable with a Gaussian distribution with zero mean and unit variance (independent of $z$) and $\epsilon$ determines the amplitude of noise; both cases of a real and a complex $\eta$ have been considered. The results are shown in Fig.~\ref{noise80}, where we compare the density profile at $t=$~60 ms with the profile at the same time once we apply a noise of intensity $\epsilon=10^{-1}$ to the initial condition. It is easy to see in this figure that the fringe pattern inside the lasing region is not washed out by the initial random noise and keeps its qualitative structure in spite of the added noise. This result is a further confirmation that the observed fringe pattern has a classical origin and that there is no need of invoking quantum phenomena to explain the experimental observations.

\begin{figure}[ht]
\begin{center}
\subfigure{\includegraphics[scale=0.85]{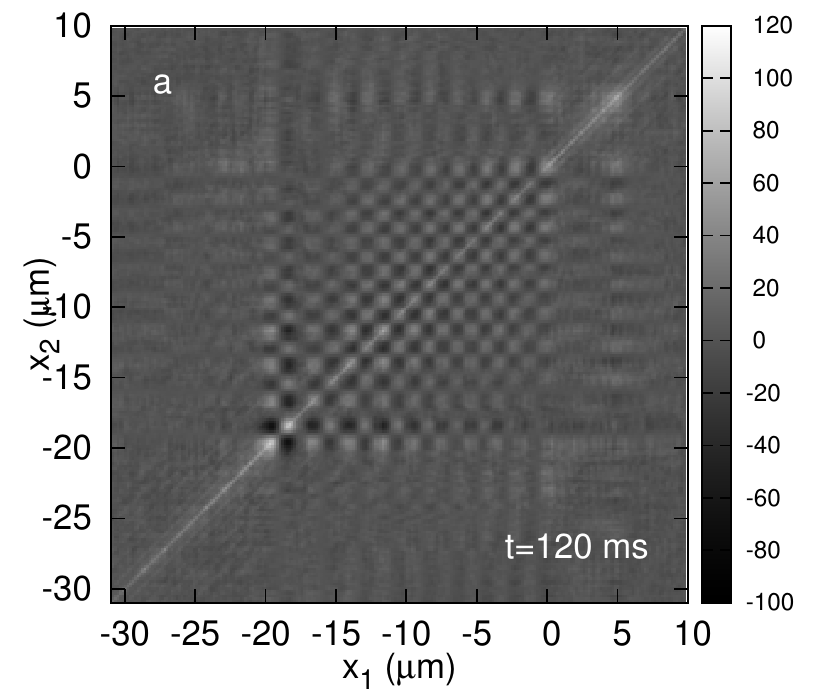}}
\subfigure{\includegraphics[scale=0.85]{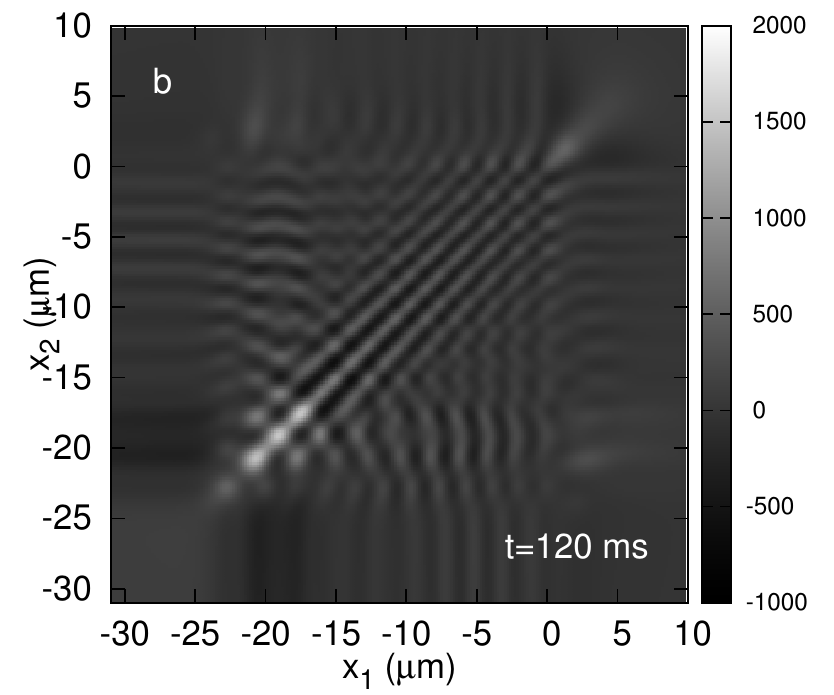}}
\end{center}
\caption{Density-density correlation function (\ref{corrfunc}) evaluated at the latest evolution time $t=$~120 ms. The left plot includes noise in the initial wavefunction according to eq.~(\ref{noiseeq}) as done in Fig.~\ref{noise80}. In the right panel, fluctuations originate from a fluctuating position of the white hole horizon as done in Fig.~\ref{densityprofiles}.}
\label{corrfig}
\end{figure}

We now proceed to explain and interpret the differences between the experimental observations and our theory that we have previously mentioned out at several points of this Letter. Inspired by a remark in the experimental article~\cite{Stein2014} that {\em ``This occurs because the position of the inner horizon} [...] {\em varies slightly from realization to realization''} we have investigated the effect of shot-to-shot fluctuations in the precise position of the white-hole horizon. Even though the strength of the optical potentials is actively stabilized to a extremely high degree in the actual experiment\footnote{J. Steinhauer, private communication.}, in the absence of more specific information on other possible mechanisms, we choose to introduce into the simulation a random gaussian noise on the height of the waterfall potential, whose main consequence is indeed (see Fig.~1c of~\cite{Stein2014}) to displace the position of the white hole horizon while keeping the black hole one fixed. We then repeated our simulation for different realizations of the random fluctuations, and we take the average of the observable data. The variance of the fluctuations is chosen in a way to optimize the qualitative agreement with the experimental observations~\cite{Stein2014}. As expected, this choice corresponds to a value of the white hole horizon displacement of the order of the fringe spacing.

The results of the average over 80 realizations of the noise (as well as over time) for the condensate and sound velocities are shown in Fig.~\ref{velocitanewnoise}. By comparing these curves with the ones in Fig.~\ref{velocitanew}, we can easily see how the contrast of the fringe pattern inside the lasing region diminishes and both curves now closely resemble the experimental ones shown in Fig.~3a of \cite{Stein2014}. 

The density profiles after different evolution times averaged over noise are shown as solid lines in Fig.~\ref{densityprofiles}. 
By comparing the dotted and the solid curves, it is apparent how the displacement of the white hole horizon greatly decreases the visibility of the fringes in the region between the two horizons: the position of the maxima and minima depends on the position of the white hole horizon and therefore changes from realization to realization, which results in destructive interference once the averages are taken. 

Our interpretation is further confirmed by the fact that the averaging procedure is more effective in washing out the fringes closer to the white hole horizon, while the ones closer to the black hole horizon are less affected. This remarkable feature is clearly visible in the simulated density profiles at late times and fully agrees with the experimental observation of Fig.~1p of~\cite{Stein2014}. For sake of completeness, we have also studied the effect of fluctuations in other quantities, e.g. the total number of particles, that do not directly affect the position of the white hole horizon, and we have verified that such effects do not produce any significant change in the observables.

\section{Correlation function}

In order to complete our analysis and confirm that our model is indeed able to reproduce all main features of the experiment, we are left with the study of the correlation function of density fluctuations, encoded in the two point function: 
\begin{equation}\label{corrfunc}
G^{(2)}(x_1,x_2)=\langle n(x_1)n(x_2) \rangle - \langle n(x_1) \rangle \langle n(x_2)\rangle \, .
\end{equation}
As done in the experiment \cite{Stein2014}, we calculate eq.~(\ref{corrfunc}) for each evolution time by taking the averages over the different realizations of the noise. The results for the latest time are shown in Fig.~\ref{corrfig} for two cases: in the left panel, we have added noise on the initial wavefunction according to eq.~\ref{noiseeq}; in the right panel, we start from a deterministic initial wavefunction but we include fluctuations in the position of the white-hole horizon as discussed above.

The left panel is characterized by a well defined chequerboard pattern in the region between the horizons, corresponding to the density modulation of the black-hole lasing mode shown in Fig.~\ref{densityprofiles}. Here, the role of the initial noise is in fact just to provide a small correction to the deterministic initial amplitude of the lasing mode.
The agreement with the experimental results in Fig.~4 of \cite{Stein2014} looks even better for the right panel where density fluctuations are due to the fluctuating position of the white hole horizon: the chequerboard pattern is present also in this case, but its clear visibility is restricted to the region close to the black hole horizon. In the vicinity of the white hole horizon around $x_{1,2}=-15\,\mu$m, it is instead partially washed out and is replaced by a series of fringes parallel to the diagonal. The bands that extend outside the lasing region into the subsonic regions beyond the horizons can be considered as a signature of the density modulation that propagates into the outer  condensate. Their vertical and horizontal orientation signals that the growing black hole lasing mode does not oscillate in time.

\section{Conclusions}

In this work we have theoretically studied the recent experiment~\cite{Stein2014} reporting the observation of self-amplifying Hawking radiation. Our work is based on a numerical solution of the Gross-Pitaevskii equation describing at mean-field level the dynamics of the condensate using the experimental parameters. As a main result of our study, we showed how such a classical approach is able to reproduce in a quantitative way the experimental observations without the need of invoking quantum fluctuation effects.

On one hand our simulations confirm the amplification mechanism, referred to as black hole laser, for an analogue black hole-white hole pair. On the other hand, they also show that the mechanism seeding the classical hydrodynamical instability responsible for the black hole laser effect has a deterministic nature, which leads to a well-defined amplitude and phase of the fringe pattern in the density profile, and can be related to a classical Bogoliubov-Cerenkov emission in the white hole horizon region. 
Finally, the fine details of the observed density profiles and of the density fluctuation correlation function can be explained in terms of classical shot-to-shot fluctuations in the position of the inner white hole horizon.

Future work will address the challenging problem of including zero-point quantum fluctuations into the theoretical model. Among the possible approaches, one may consider generalizing the Wigner method~\cite{wigner} from single black hole horizons in infinitely extended condensates~\cite{Carusotto2008} to more realistic geometries. If successful, this study will have a crucial importance to characterize in a quantitatively accurate way the observable consequences of spontaneous Hawking processes. Intriguing experimental data in this direction were recently published in the preprint \cite{stein-latest} and are presently under active scrutiny by the analogue model community. During review of this work, a related analysis of the experiment in [1] has appeared~\cite{jacobson}. \\

We are grateful to R. Parentani and J. Steinhauer for continuous instructive discussions and a critical reading of the manuscript. M.T. gratefully thanks R. Parentani for the kind hospitality in Paris. I.C.'s activity was supported by the ERC through the QGBE grant, by the EU-FET Proactive grant AQuS, Project No. 640800, and by the Autonomous Province of Trento, partially through the project ``On silicon chip quantum optics for quantum computing and secure communications'' (``SiQuro'').

\bibliographystyle{ieeetr}
\bibliography{Bibliography}

%
%

\end{document}